\documentclass[conference]{IEEEtran}
\usepackage{graphicx,color,epsfig}
\usepackage{fancyhdr,fancybox}
\usepackage{varioref,float,amssymb,amsmath}
\usepackage{float,amsfonts,amsthm}
\usepackage{changebar,longtable}
\usepackage[latin1]{inputenc}
\usepackage{fancyheadings}
\usepackage{makeidx}
\usepackage{psfrag}
\usepackage[below]{placeins}
\usepackage{flafter}
\usepackage[small]{caption2}

\newcommand{\bra}[1]{\left\langle{#1}\right\vert}
\newcommand{\ket}[1]{\left\vert{#1}\right\rangle}
\newcommand{\xxx}[1]{{#1}}%
\begin{document}

\title{Quantum Search Algorithms on \\ Hierarchical Networks}

\author{\IEEEauthorblockN{Franklin de Lima Marquezino}
\IEEEauthorblockA{Universidade Federal do Rio de Janeiro\\
Rio de Janeiro, RJ 21941-972, Brazil \\
Email: franklin@cos.ufrj.br}
\and
\IEEEauthorblockN{Renato Portugal}
\IEEEauthorblockA{Laborat\'{o}rio Nacional de Computa\c{c}\~{a}o Cient\'{\i}fica\\
Petr\'opolis, RJ 25651-075, Brazil \\
Email: portugal@lncc.br}
\and
\IEEEauthorblockN{Stefan Boettcher}
\IEEEauthorblockA{Emory University\\
Atlanta, GA 30322-2430, USA\\
Email: sboettc@emory.edu
}}

\maketitle

\begin{abstract}
The ``abstract search algorithm'' is a well known quantum method to find a marked vertex in a graph. It has been applied with success \xxx{to} searching algorithms for the hypercube and the two-dimensional grid. In this work we provide an example for which that method fails to provide the best algorithm in terms of time complexity. We analyze search algorithms in degree-3 hierarchical networks using quantum walks driven by non-groverian coins. Our conclusions are based on numerical simulations, but the hierarchical structures of the graphs seems to allow analytical results.
\end{abstract}

\section{Introduction}

Searching algorithms play an important role in quantum computing. One
of the \xxx{best-}known quantum algorithms is \xxx{due to} Grover~\cite{Gro97a}, which allows one to find a marked item in an unsorted database quadratically faster compared with the best classical algorithm. It uses an important technique called amplitude amplification, which can be applied in many computational problems providing gain in time complexity.

A related problem is to find a marked location \xxx{in} a spatial,
physical region. Benioff~\cite{Ben02} asked how many step\xxx{s are}
necessary for a quantum robot to find a marked \xxx{vertex} in a
two-dimensional grid with $N$ \xxx{vertices}. In his model, the robot can move
from one \xxx{vertex} to an adjacent one spending one time unit. Benioff
showed that a direct application of \xxx{the} Grover algorithm does not provide improvements in the time complexity compared to a classical robot, which is $O(N)$. Using a different technique, called \textit{abstract search algorithms}, Ambainis et.~al.~\cite{AKR05} showed that it is possible to find the marked \xxx{vertex} with $O\left(\sqrt{N}\log N\right)$ steps. Tulsi~\cite{Tul08} was able to improve this algorithm obtaining the time complexity $O\left(\sqrt{N\log N}\right)$.

The time needed to find a marked \xxx{vertex} depends on the spatial
layout. The abstract search algorithm is a technique that can be
applied to any regular graph. It is based on a modification of the
standard discrete quantum walk. The coin is the Grover operator for
all \xxx{vertices} except for the marked one which is $-I$. The \xxx{choice of
  the} initial condition is also \xxx{essential}. It must be the uniform superposition of all states of the computational basis of the coin-position space. This technique was applied with success \xxx{on} higher dimensional grids~\cite{AKR05}, honeycomb networks~\cite{ADMP10}, regular lattices~\cite{HT10} and triangular networks~\cite{ADFP11}.

In this work, we analyze a hierarchical network called Hanoi network (HN3)~\cite{BGA08}, for which the abstract search algorithm does not provide the most efficient search algorithm. HN3 is a special case of \textit{small world networks}, which are being used in many contexts including quantum computing~\cite{GGS05,MPB07}. In order to obtain a quicker algorithm, we modify the coin operator and the initial condition used in the abstract search algorithm  to take advantage of the small world structure. Our results are based on numerical simulations, but the hierarchical structure of HN3 indicates that analytical results can also be obtained.

The structure of the paper is as follows. Sec.~\ref{sec:HS}
\xxx{introduces the} degree-3 Hanoi network. Sec.~\ref{sec:QW} describes the standard coined discrete quantum walk on HN3. Sec.~\ref{sec:ASA} reviews the basics of the abstract search algorithm. Sec.~\ref{sec:MM} describes the modifications we propose to enhance the time complexity of quantum search algorithms on HN3. Sec.~\ref{sec:Results} describes the main results based on numerical simulations. Finally, we present our conclusion in Sec.~\ref{sec:Conclusions}.

\section{Hierarchical Structures}\label{sec:HS}
\begin{figure}[h]
    \centering
    \includegraphics[height=5.0cm]{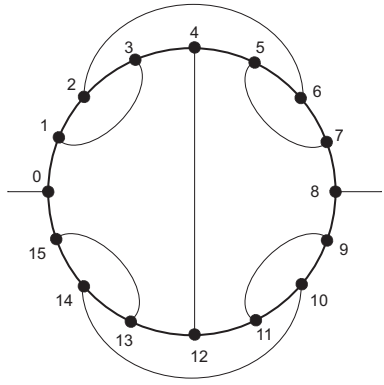}
    \caption{HN3 with 16 vertices.}
    \label{fig:NH3}
\end{figure}

The Hanoi network has a cycle with $N=2^n$ vertices as a backbone
structure, that is, each vertex is adjacent to 2 neighboring vertices
in this structure and extra, long-range edges are introduced with the
goal of obtaining a small-world hierarchy. The labels of the vertices
$0< k \le 2^n-1$ can be factorized as
\begin{equation}\label{eq:k}
    k =2^{k_1} (2\,{k_2} + 1),
\end{equation}
where ${k_1}$ denotes the level in the hierarchy and ${k_2}$ labels
consecutive vertices within each hierarchy. In any level, one links
the vertices with consecutive values of ${k_2}$ keeping the degree
constraint. When ${k_1}=0$, the values of $k$ are the odd
integers. For HN3, we link 1 to 3, 5 to 7 and so on. The \xxx{vertex} with
label 0, not being covered by Eq.~(\ref{eq:k}), is linked to the \xxx{vertex}
of label $2^{n-1}$. Fig.~\ref{fig:NH3} shows all edges for HN3 when
the number of vertices is $16$.  Using this figure, one can easily
built HN3 \xxx{recursively, each time doubling the number of} vertices. Our analysis \xxx{will} be performed for a generic value of $n$ to allow us to determine the computational cost as function of $N$.

HN3 has a small-world structure because the diameter of the
\xxx{network only increases with $\sim\sqrt{N}$~\cite{BGA08},  less fast than the
  number of vertices $N$. Yet, HN3 is a regular graph of fixed degree
  $d=3$ at each vertex.}

\section{Quantum Walks on Hierarchical Structures}\label{sec:QW}

A coined quantum walk in HN3 with $N=2^n$ vertices has a Hilbert space ${\cal H}_C\otimes {\cal H}_P$, where ${\cal H}_C$ is the $3$-dimensional coin subspace and ${\cal H}_P$ the $N$-dimensional position subspace. A basis for ${\cal H}_C$ is the set $\{\ket{a}\}$ for  $0\leq a\leq 2$ and ${\cal H}_P$ is spanned by the set $\{\ket{k}\}$ with, $0 \leq k\leq N-1$. We use the decomposition $k=(k_1,k_2)$, given by Eq.~(\ref{eq:k}), when convenient. A generic state of the discrete quantum walker in HN3 is
\begin{equation}
\ket{\Psi(t)}=\sum_{a=0}^{2}\sum_{k=0}^{N-1}\psi_{a,k}(t)\ket{a}\ket{k}. \label{eq:estgeral1d}
\end{equation}

The evolution operator for the standard quantum walk~\cite{Kem03} is
\begin{equation}\label{evol}
U=S\circ(C\otimes I),
\end{equation}
where $I$ is the identity in ${\cal H}_P$ and $S$ is the shift operator defined by
\begin{eqnarray*}
  S\ket{0}\ket{k_1,k_2} &=& \ket{0}\ket{k_1,k_2+(-1)^{k_2}},
\end{eqnarray*}
and
\begin{eqnarray*}
  S\ket{1}\ket{k} &=& \ket{2}\ket{k+1}, \\
  S\ket{2}\ket{k} &=& \ket{1}\ket{k-1}.
\end{eqnarray*}
The arithmetical operations on the second ket is performed modulo $N$.
The shift operator obeys $S^2=I$.  $C$ is a unitary coin operation in ${\cal H}_C$. In the standard walk, $C$ is the Grover coin, denoted by $G$,
\begin{equation}
    G=\frac{1}{3}\left[ \begin {array}{ccc} -1&2&2\\\noalign{\medskip}2&-1&2
\\\noalign{\medskip}2&2&-1\end {array} \right],
\end{equation}
which is the most diffusive coin~\cite{NV00}.

For example, using Figure~\ref{fig:NH3} we see that
\begin{eqnarray*}
  S\ket{0,2} &=& \ket{0,6} \\
  S\ket{1,4} &=& \ket{2,5} \\
  S\ket{2,5} &=& \ket{1,4}
\end{eqnarray*}
If the value of the coin is zero, the walker takes the edge that leaves the main circle. If the value of the coin is 1, the walker goes clockwise and inverts the value of the coin $(1\rightarrow 2)$. If the value of the coin is 2, the walker goes counterclockwise and inverts the value of the coin $(2\rightarrow 1)$.

The dynamics of the standard quantum walk is given by
\begin{equation}\label{eq:U_t}
   \ket{\Psi(t)}= U^t \ket{\Psi_0},
\end{equation}
where $\ket{\Psi_0}$ is the initial condition. After $t$ steps of unitary evolution, we perform a position measurement which yields a probability distribution given by
\begin{equation}\label{eq:PROB  }
    p_k = \sum_{a=0}^2 \left| \bra{a,k}U^t \ket{\Psi_0} \right|^2.
\end{equation}

\section{Abstract Search Algorithms}\label{sec:ASA}

The \textit{abstract search algorithm}~\cite{AKR05} is based on a modified evolution operator $U^\prime=S \cdot C^\prime$, obtained from the standard quantum walk operator $U$ by replacing the coin operation $C$ with a new unitary operation $C^\prime$ which is not restricted to ${\cal H}_C$ and acts differently on the searched vertex. The modified coin operator is
\begin{equation}\label{eq:Cprime}
    C'= -I\otimes \ket{k_0}\bra{k_0} + C\otimes (I- \ket{k_0}\bra{k_0}),
\end{equation}
where $k_0$ is the marked vertex in a regular graph and $C$ is the Grover coin $G$, the dimension of which depends on the degree of the graph. Ambainis et.~al.~\cite{AKR05} have shown that the time complexity of the spatial search algorithm can be obtained from the spectral decomposition of the evolution operator $U$ of the unmodified quantum walk, which is usually simpler than that of $U'$.

The initial condition $\ket{\psi_0}$ is the uniform superposition of all states of the computational basis of the whole Hilbert space. This can be written as the tensor product of the uniform superposition of the computational basis of the coin space with the uniform superposition of the position space. Usually, this initial condition can be obtained in time $O(\sqrt N)$, where $N$ is the number of vertices.

The evolution operator is applied recursively starting with the initial condition $\ket{\psi_0}$. If $t_f$ is the running time of the algorithm, the state of the system just before measurement is $U'^{t_f}\ket{\psi_0}$. If one analyzes the probability of obtaining the marked vertex $k_0$ as function of time since the beginning of the algorithm, one gets an oscillatory function with the first maximum at $t_f$.

\section{Modified Method}\label{sec:MM}

The coin in a quantum walk is used to determine the direction of the
movement. \xxx{The }Grover coin is an isotropic operator regarding all outgoing
edges from a vertex. It is useful in networks that have no special
directions, such as two-dimensional grids and hypercubes. The Hanoi
network, on the other hand, has a special direction that creates the
small world structure. Any edge that takes the walker outside the
circular backbone provides an interesting opportunity in terms of
searching. The strategy is to have a parameter that can control the
probability flux among the edges, reinforcing or decreasing the flux
outwards or inwards the circular backbone.

Instead of using \xxx{the} Grover coin of the \textit{abstract search
  algorithms}, we use
\begin{equation}\label{eq:new_C}
  \begin{split}
    C = \frac{2\epsilon}{d}\ket{0}\bra{0} + 2\sqrt{\frac{\epsilon(d-\epsilon)}{d^2(d-1)}}\sum_{j=1}^{d-1}\left(\ket{j}\bra{0} + \ket{0}\bra{j}\right) +\\
    \frac{2(d-\epsilon)}{d^(d-1)}\sum_{j,j'=1}^{d-1}\left(\ket{j}\bra{j'}
      + \ket{0}\bra{j}\right) - I,
  \end{split}
\end{equation}
\xxx{where $d=3$ is the degree at each vertex.}
When $\epsilon=1$, \xxx{the} Grover coin is recovered. When $0<\epsilon<1$, the
probability flux  \xxx{along small-world edges} which escapes from the
circular backbone is weakened.  When $1<\epsilon<3$, the probability
flux  \xxx{off the backbone} is reinforced, allowing the walker to
use the small world structure with higher efficiency. \xxx{Hence,}
this new coin \xxx{controls the bias to escape off the circular backbone
  of HN3 through the parameter $\epsilon$.}

The \textit{abstract search algorithms} use a uniform distribution as
initial condition. We change this recipe. The initial condition is
\begin{equation}\label{eq:initial_condition}
  \ket{\psi(0)}=\sqrt{\frac{\epsilon}{d}}\ket{0}\ket{s} + \sqrt{\frac{d-\epsilon}{d(d-1)}}\sum_{j=1}^{d-1}\ket{j}\ket{s},
\end{equation}
where $\ket{s}$ is the uniform superposition on the position
space. When $\epsilon=1$ the initial condition is the uniform
superposition of coin-position space.

The analysis of the evolution of the quantum search algorithm using
the new coin and initial condition is far more complex than the
standard one. Our conclusions \xxx{here} are based in numerical
simulations. Since the standard abstract search algorithm with Grover
coin is obtained with $\epsilon=1$, our simulations allow us to
compare the cost of the new search algorithm with the standard one and
to determine the value of $\epsilon$ that optimizes the total cost of
the algorithm.

\section{Results}\label{sec:Results}

\begin{figure}[!h]
  \centering
  \includegraphics[width=0.9\columnwidth]{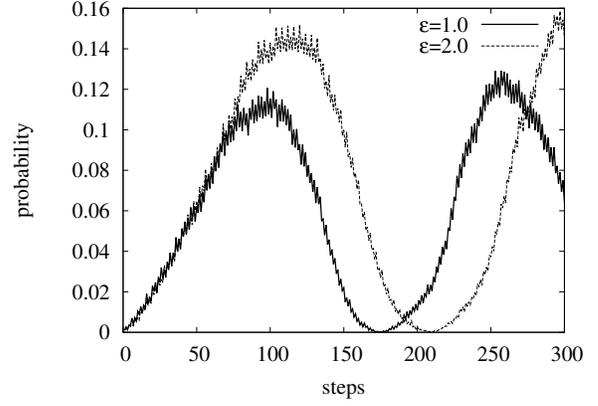}
  \caption{Probability of finding walker at vertex $k_0$ as a function
    of time for two values of $\epsilon$.}
  \label{fig:prob-vs-t}
\end{figure}

Fig.~\ref{fig:prob-vs-t} shows the oscillatory behavior of the
probability of finding the walker at the marked vertex
$k_0$. Initially, the probability is close to zero, because the
initial condition is a state that is close to the uniform
superposition of all vertices. The running time of the algorithm is
the value of $t$ for which the probability reaches its first
maximum. Note that for $\epsilon=1$, which is the \textit{abstract
  search algorithm}, the maximum value of the probability is smaller
than that of $\epsilon=2$. The maximum value for $\epsilon=2$ occurs
at a time latter than that of $\epsilon=1$. This is not a problem, as
we show when we analyze the total cost of the algorithm. \xxx{In
  either case}, the
maximum value of the probability is not close to 1, as one would
expect in order to have high probability to find the marked
vertex. This means that the algorithm must be rerun many times to
amplify the success probability.

\begin{figure}[!h]
  \centering
  \includegraphics[width=0.9\columnwidth]{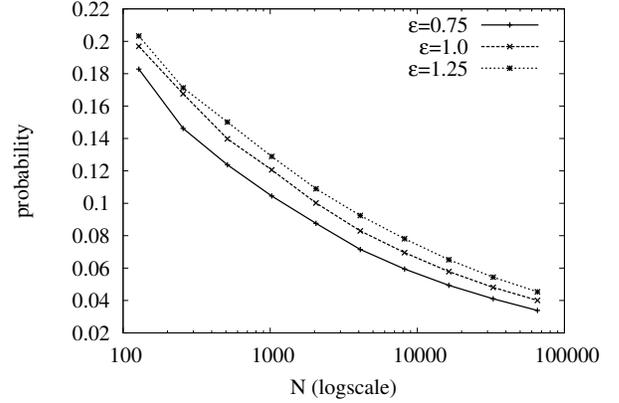}
  \caption{Success probability of the search algorithm as a function
    of $N$ for three values of $\epsilon$.}
  \label{fig:prob-vs-N}
\end{figure}

Fig.~\ref{fig:prob-vs-N} shows the success probability\xxx{, defined
  as the first peak in Fig.~\ref{fig:prob-vs-t},} of a single run
of the search algorithm as a function of the network size $N$ for
different values of $\epsilon$. The horizontal axis is in logscale for
better visualization.  From the line inclination, we conclude that the
success probability decays approximately as $O(1/\log N)$. It means
that the algorithm must be rerun around $O(\log N)$ times in order to
ensure a success probability close to $1$.

\begin{figure}[!h]
  \centering
  \includegraphics[width=0.9\columnwidth]{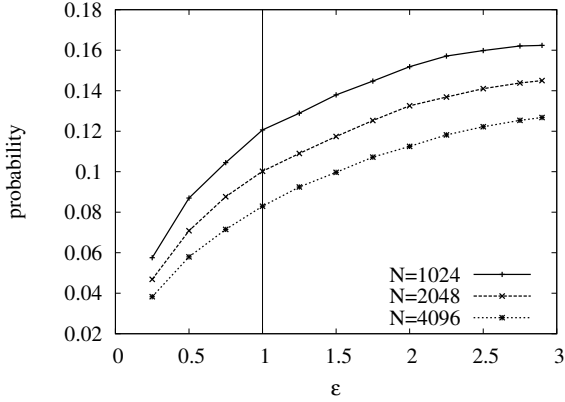}
  \caption{Success probability as function of parameter $\epsilon$ for
    three values of $N$.}
  \label{fig:prob-vs-epsilon}
\end{figure}

Fig.~\ref{fig:prob-vs-epsilon} shows the success probability as
function of parameter $\epsilon$ for three values of $N$. The success
probability is an increasing function in terms of $\epsilon$. The
range $\epsilon<1$ is the worst one. This can be explained in the
following way. For small values of $\epsilon$, the walker is
constrained to live in the HN3 backbone cycle. It is known that
quantum and classical search algorithms in cycles have the same time
complexity. On the other hand, in the range $\epsilon>1$, the success
probability is greater than the one obtained in the \textit{abstract
  search algorithm}. Recall that when $\epsilon>1$, the probability
flux  \xxx{along small-world edges}  is reinforced. This is a strong
indication that small world structures improve quantum search
algorithms.

\begin{figure}[!h]
  \centering
  \includegraphics[width=0.9\columnwidth]{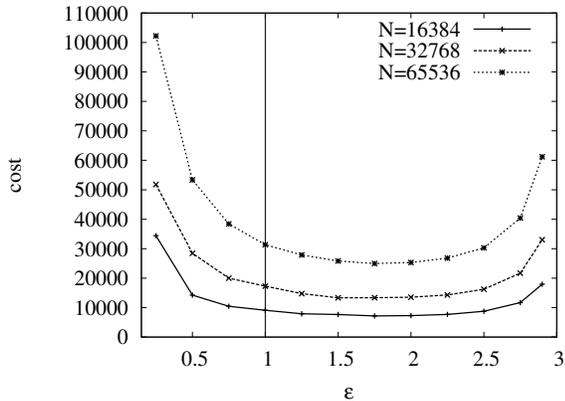}
  \caption{Computational cost of the search algorithm in terms of
    $\epsilon$ for three values of $N$.}
  \label{fig:cost-vs-epsilon}
\end{figure}

The  \xxx{perspective provided by} Fig.~\ref{fig:prob-vs-epsilon} is
\xxx{incomplete for HN3, though. While the highest values of
$\epsilon$ appear to be favorable,} the analysis of the total cost of
the algorithm points out to another value.
Fig.~\ref{fig:cost-vs-epsilon} shows the computational cost of the
search algorithm in terms of $\epsilon$ for three values of $N$. The
minimum cost occurs for $\epsilon\approx1.7$, which is \xxx{well
  above the value for the} Grover coin, $\epsilon=1$. This clearly shows that
the \textit{abstract search algorithm} does not provide the best
algorithm for searching a marked vertex in the hierarchical
structures. \xxx{In turn, minimizing cost occurs also well below the
  value of $\epsilon$ that maximized probability. This discrepancy
 likely  originates with the ``confinement'' effect already observed
 for a classical walker on HN3~\cite{BGA08}: To utilize ever-longer
 small-world edges off the backbone, a walker has to be able to mover ever further
 along the backbone between those edges. At exactly $\epsilon=3$, the
 probability for moving along the backbone vanishes, and a walker
 remains confined to its present small-world edge and the cost in
 Fig.~\ref{fig:cost-vs-epsilon} diverges.}

\xxx{Naive analogy with the exact renormalization group (RG)
  analysis for the classical walker~\cite{BGA08} would suggest that
  the cost of a quantum walk would be minimized in a boundary layer for $\epsilon\to3$
  with a width that decreases as some power of the system size $N$.
Our simulations in Fig.~\ref{fig:cost-vs-epsilon} exhibit a very broad
cost-minimum near $\epsilon\approx1.7$, especially for small
values of $N$. For the system  size attained in our simulations, it is
difficult to decide whether the minimum point will converge to a
finite value of $\epsilon<3$ or approach $\epsilon\to3$ for large
$N$.}
We are currently
analyzing search algorithms \xxx{on another Hanoi network,
  HN4~\cite{BGA08}, which is 4-regular and does not suffer
  confinement. While otherwise identical to HN3, in HN4 two small-world edges extend
  off the backbone
  symmetrically in \emph{both} directions from each vertex.}

\begin{figure}[!h]
  \centering
  \includegraphics[width=0.9\columnwidth]{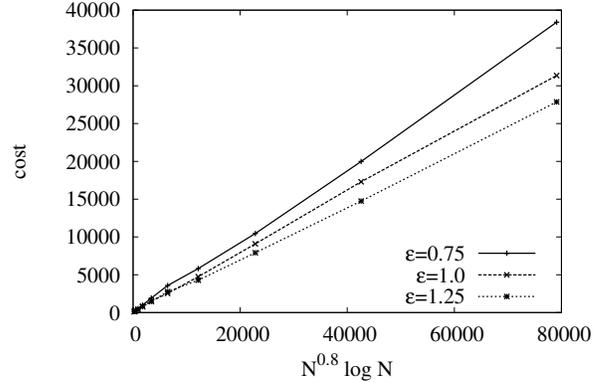}
  \caption{Computational cost of the search algorithm in terms of
    $N^{0.8}\log N$ for three values of $\epsilon$.}
  \label{fig:cost-vs-N}
\end{figure}

Fig.~\ref{fig:cost-vs-N} shows the computational cost of the search
algorithm as a function of the network size $N$ for different values
of $\epsilon$.  Those plots complement the plot of
Fig.~\ref{fig:cost-vs-epsilon}.  From the line inclination, we
conclude that the cost is $O(N^c \log N)$, where $c$ is a constant
close to $0.8$. It is clear from the plots that the curve
corresponding to the lowest time complexity has $\epsilon > 1$,
outperforming the Grover coin.

\section{Conclusion}\label{sec:Conclusions}

The \textit{abstract search algorithm} is a powerful technique for the
development and analysis of quantum search algorithms on regular
graphs.  The quantum speed-up achieved by the abstract search
algorithm strongly depends on the properties of the graph under
consideration. In this work, we study the quantum search on the Hanoi
network (HN3). This particular graph presents a sufficiently intricate
structure to exhibit not-trivial properties for statistical models,
yet sufficiently simple to reveal analytical insights.

We proposed a modification on the abstract search algorithm for HN3
and numerically analyzed its performance. Our modification consisted
on changing the coin operator in order to have a parameter $\epsilon$
that \xxx{ controls the probability flux among the edges, enhanced either
along or off   the circular backbone.} Our
modified method also uses a parameterized initial condition instead of
the uniform distribution.

Our numerical results \xxx{clearly}  show that the performance of the search
algorithm is better when we choose the parameter $\epsilon>1$ to
reinforce the probability flux  \xxx{along small-world edges} instead of
the circular backbone (which would correspond to $\epsilon<1$). The
success probability of a single run of the search algorithm is
considerably higher whenever we choose $\epsilon>1$. The total cost
of the algorithm \xxx{appears to reach a minimum for
  $\epsilon\approx1.7$. The cost appears to scale with
  $\sim N^{0.8}$ with the system size.  A more elaborate
 analysis used for the classical walker in Ref.~\cite{BGA08}
  and for quantum transport through HN3 in Ref.~\cite{BVN11} is
  required to provide exact results.}

We conclude that the abstract search algorithm, which uses the Grover
coin to drive the quantum walk on a regular graph, does not provide
the most efficient search algorithm for the Hanoi network HN3.

Our works on progress are now focused on the analysis of search
algorithms for the Hanoi network HN4.

\section*{Acknowledgment}

The authors would like to thank Dr. Reinaldo Rosa for helping
to start this collaboration.

\end{document}